\documentclass[namedreferences]{solarphysics}

\usepackage[hyperref,optionalrh]{spr-sola-addons} % For Solar Physics 

\hypersetup{
    colorlinks=true,
    allcolors=blue,
}

\usepackage{graphicx}        % For eps figures, newer & more powerful
\usepackage{color}           % For color text: \color command
\usepackage{breakurl}        % For breaking URLs easily trough lines
\usepackage{tabto}
\usepackage{gensymb}
\usepackage{enumerate}% http://ctan.org/pkg/enumerate

\usepackage{tikz}

\usepackage[detect-none]{siunitx}
\renewcommand{\vec}[1]{{\mathbfit #1}}

\newcommand{\bb}{\vec B}

\chardef\us=`\_

\begin{document}

\begin{article}
\begin{opening}

\title{Estimating Total Open Heliospheric Magnetic Flux}

\author[addressref={aff1,aff2},corref,email={samantha.wallace@nasa.gov}]{\inits{S.}\fnm{S.}
\lnm{Wallace}{\orcid{0000-0002-1091-4688}}}%\sep

\author[addressref=aff2]{\inits{C.N.}\fnm{C.N.}~\lnm{Arge}}%\sep

\author[addressref=aff3]{\inits{M.}\fnm{M.}~\lnm{Pattichis}}%\sep

\author[addressref={aff4},corref]{\inits{R.A.}\fnm{R.A.}~\lnm{Hock-Mysliwiec}}%\sep

\author[addressref={aff4},corref]{\inits{C.J.}\fnm{C.J.}~\lnm{Henney}}

%\author{\inits{}\fnm{}~\lnm{}\orcid{}}
%\author{P.~\surname{Author-a}$^{1}$\sep
%        E.~\surname{Author-b}$^{1}$\sep
%        M.~\surname{Author-c}$^{2}$      
%       }

%   \institute{$^{1}$ First affiliation
%                     email: \url{e.mail-a} email: \url{e.mail-b}\\ 
%              $^{2}$ Second affiliation
%                     email: \url{e.mail-c} \\
%             }

\address[id=aff1]{Department of Physics and Astronomy, University of New Mexico 1919 Lomas Blvd. NE, Albuquerque, NM, 87131, USA}
\address[id=aff2]{Heliophysics Science Division, NASA Goddard Space Flight Center, Code 671, Greenbelt, MD, 20771, USA}
\address[id=aff3]{Department of Electrical and Computer Engineering, University of New Mexico 498 Terrace St NE, Albuquerque, NM 87106, USA}
\address[id=aff4]{Air Force Research Laboratory/Space Vehicles Directorate, Kirtland Air Force Base, NM, 87117, USA}

\runningauthor{S. Wallace {\it et al.}}
\runningtitle{Estimating Total Open Heliospheric Flux}
%\runningtitle{Comparing Methods for Estimating Total Open Flux}

\begin{abstract}

Over the solar activity cycle, there are extended periods where significant discrepancies occur between the spacecraft-observed total (unsigned) open magnetic flux and that determined from coronal models.  In this article, the total open heliospheric magnetic flux is computed using two different methods and then compared with results obtained from {\it in-situ} interplanetary magnetic field observations. The first method uses two different types of photospheric magnetic-field maps as input to the Wang--Sheeley--Arge (WSA) model: (i) traditional Carrington or diachronic maps, and (ii) Air Force Data Assimilative Photospheric Flux Transport model synchronic maps.  The second method uses observationally derived helium and extreme ultraviolet coronal hole maps overlaid on the same magnetic field maps in order to compute total open magnetic flux. The diachronic and synchronic maps are both constructed using magnetograms from the same source, namely the National Solar Observatory Kitt Peak Vacuum Telescope and Vector Spectromagnetograph. The results of this work show that total open flux obtained from observationally derived coronal holes agrees remarkably well with that derived from WSA, especially near solar minimum. This suggests that, on average, coronal models capture well the observed large-scale coronal hole structure over most of the solar cycle. Both methods show considerable deviations from total open flux deduced from spacecraft data, especially near solar maximum, pointing to something other than poorly determined coronal hole area specification as the source of these discrepancies. 

\end{abstract}
\keywords{Magnetic fields, Interplanetary; Coronal holes; Corona, Models}
\end{opening}
%-------------------------------------------------

%%%%%%%%%%%%%%%%%%%%%%%%%%%%%%%%%%%%%%%%%%%%%%%%%%%%%%%%%%%%%%%%%
%%%%%%%%%%%%%%%%%%%     Begin Section 1   %%%%%%%%%%%%%%%%%%%%%%%
%%%%%%%%%%%%%%%%   Introduction / Background  %%%%%%%%%%%%%%%%%%%
%%%%%%%%%%%%%%%%%%%%%%%%%%%%%%%%%%%%%%%%%%%%%%%%%%%%%%%%%%%%%%%%%

\section{Introduction}
     \label{1-Intro} 
The photosphere is permeated by magnetic fields of varying strengths and spatial scales that form topologies described as either closed or open. While closed magnetic field lines can be described as traceable between photospheric footpoints of opposite polarity along paths not extending beyond the corona, this article concerns the Sun's unipolar open magnetic field lines that extend out into the heliosphere forming the interplanetary magnetic field (IMF). These ``open'' magnetic fields, which remain connected to the photosphere at one end and yet must reconnect within the interstellar medium, play a critical role in space weather and geomagnetic activity.  The primary sources of open heliospheric magnetic flux and the fast solar wind are coronal holes.  Coronal holes are mostly unipolar regions that are observed to have lower density and temperature relative to the background corona resulting in reduced coronal emission, appearing dark in X-ray and extreme ultraviolet (EUV) images \citep[see][and references therein]{Cranmer2009}. However, coronal hole regions are observed to be brighter in the helium (He) {\sc I} 10830 \AA\ chromospheric line due to decreased absorption at these wavelengths  \citep{HarveyRecely2002}.\\
\tab \hspace{5mm}Total open heliospheric flux is determined using various modeling and observational methods. Simple magnetostatic potential field source surface (PFSS) models \citep[\textit{e.g.}][]{Shatten1969,Altschuler1969,WangSheeley1992} as well as advanced magnetohydrodynamic (MHD) coronal models \citep[\textit{e.g.}][and references therein]{Lionello2009}, are routinely used to estimate coronal hole open flux areas. Both models rely on input global photospheric-field maps assembled from full-disk observations of the line-of-sight photospheric field to derive the coronal field.  These input maps are constructed in a variety of ways (discussed further in Section~\ref{2-ModelsVSobs}), representing either an entire Carrington rotation or one moment in time.\\  
\tab \hspace{5mm}During the Ulysses solar minimum and maximum fast latitude scans, \citet{Smith2001} discovered the latitude invariance of $R^{2}|B_\text{r}|$, which permits total unsigned heliospheric flux, $4{\pi}R^{2}|B_\text{r}|$, to be estimated from single point measurements, where {\it R} is the heliocentric distance and $|B_\text{r}|$ is the unsigned radial component of the heliospheric magnetic field.  While this remarkable result was met with some initial skepticism \citep[\textit{e.g.}][]{Arge2002}, a study by \citet{Lockwood2004} tested the validity of deriving open flux under this assumption using observations from near-Earth spacecraft and Ulysses. They concluded that measurements of the radial field can be used to derive the total open heliospheric flux with $\approx$5\,\% uncertainty or less if averaging over 27 days or more.\\   
\tab \hspace{5mm} Discrepancies between the total open flux estimates derived from models and {\it in-situ} observations have been noted for a while, with spacecraft generally yielding larger open flux estimates than models.  \citet{Owens2008a} found that {\it in-situ} measurements from well-separated heliospheric spacecraft inside 2.5 AU give self-consistent results of total open heliospheric flux.  However, they found that open flux estimates from multiple spacecraft at the same \textit{R} increasingly vary as \textit{R} increases, with the largest discrepancies occurring beyond 2.5 AU (\citealt{Owens2008a}, Figure 1).  They suggested that this could be attributed to a reduced signal-to-noise ratio in spacecraft detection of $B_\text{r}$ further away from the Sun due to tangential and meridional fluctuations along each magnetic field line that result in these field components bleeding into $B_\text{r}$ measurements \citep{Owens2008a}. In a later set of articles, \citet{Lockwood2009a} and \citet{Lockwood2009b,Lockwood2009c} revisited this topic and attribute what they term the ``flux excess effect'' not to small-scale structure introduced by the field line's own propagation, but to separate large-scale longitudinal solar wind structures that distort the magnetic field (\citealp{Lockwood2009c}, Figure 1).  They derived a correction term to apply to spacecraft $B_\text{r}$ measurements to account for these kinematic effects, and when this is applied, for the time period investigated in their study they produce good agreement between PFSS and spacecraft-derived open flux (\citealp{Lockwood2009c}, Figure 9).  \citet{Owens2017} accounted for this same kinematic contribution to $B_\text{r}$ instead using observations of sunward suprathermal electron beam or ``strahl'' to filter out instances where field lines were locally inverted at 1 AU.  Their method resulted in higher open fluxes than the kinematic correction, closer to results obtained when averaging $B_\text{r}$ over one day before taking the unsigned value.  Both methods on average fell within the results of averaging $B_\text{r}$ between one to five days before taking the unsigned value (\citealp{Owens2017}, Figure 5); however, unlike the temporal averaging technique, these methods attempt to account for the flux excess with more physical justification.\\ 
\tab \hspace{5mm} Another way that the differences between both model-derived and spacecraft-observed open flux have been explained is by applying correction factors to the photospheric-field maps to drive coronal models. \citet{Wang2000} compared estimates of total open flux from 1971\,--\,1998 using both the PFSS model and spacecraft observations and found, with a few notable exceptions ({\it e.g.}
1978\,--\,1980 and 1985\,--\,1989), that they agree remarkably well when using a latitude [$\lambda$] dependent saturation correction factor.  The photospheric field used to derive the open flux was a combination of traditional Carrington photospheric magnetic- field synoptic maps from Mount Wilson (MWO) and Wilcox (WSO) Solar Observatories.  It was necessary to use WSO maps from 1976\,--\,1995 because the MWO magnetograph suffered from instrumentation issues during this period. They also assumed, based on previous work \citep{WangSheeley1995}, that the MWO correction factor (4.5\,--\,2.5$\sin^2\lambda$) was equally applicable to both the MWO and WSO observations, as both the instruments are very similar to each other and use the same Fe I 5250 \AA\ line to derive the magnetic field.  However, \citet{Svalgaard2006} challenged the application of the MWO saturation correction to WSO data, arguing that it is specific to the MWO instrument and that a constant correction factor of 1.85 (updated from \citealt{Svalgaard1978}, originally 1.8) is the appropriate correction factor for WSO data.  Other studies, spanning several years (\citealp{Owens2008b}, \citealp{Linker2016}) and focusing on one Carrington rotation \citep{Linker2017}, used photospheric-field maps derived from other observatories and found similar discrepancies between model-derived and spacecraft-measured $|B_\text{r}|$. In all of these studies, {\it in-situ} observations on average produced larger $|B_\text{r}|$ estimates than models, except \citet{Linker2016} where a correction factor of 1.5 was applied to the photospheric-field maps for the results to agree.\\ 
\tab \hspace{5mm} \citet{Riley2007} separately applied both a constant 1.8 scaling factor and the MWO latitude-dependent correction to WSO input maps from 1976\,--\,2007 and calculated total open flux using the PFSS model for both sets of maps. He found that the maps with the MWO saturation correction factor applied to them produced the best overall agreement with the heliospheric derived fluxes but noted significant mismatches during the periods of solar minimum, exactly when one would expect the agreement to be best. For the case of the 1.8 saturation factor applied to the WSO maps, the open-flux results agreed well with the heliospheric values only during solar minimum and under-predicted them for all other times.  To explain this Riley proposed and tested the idea that interplanetary coronal mass ejections (ICMEs) propagating out into space yet still magnetically connected back to the Sun are detected by spacecraft in the same way that coronal hole open flux is. Riley argued that the total unsigned heliospheric flux is comprised of open coronal hole and closed ICME fluxes, especially during solar maximum.  This assumption is reasonable, since distinguishing between open field lines and closed ones from ICMEs still rooted back at the Sun in spacecraft observations, especially when the front of the ICME is far beyond 1 AU, is generally very difficult. Using a simple model to estimate heliospheric ICME flux from sunspot number, Riley found an extremely good match between the combined ICME plus WSO open-flux estimates and heliospheric values. While not claiming to have settled the issue, Riley's results nonetheless provided an alternative explanation for the temporal variation of observed heliospheric flux without having to resort to correcting the WSO magnetic field data with the MWO saturation correction factor.  Owens et al., in a series of articles  \citep{Owens2006, Owens2007, Owens2011}, similarly argued that much, if not all, of the variation of heliospheric flux can be explained by a constant open solar magnetic flux baseline or floor with the addition of a time-varying, solar-cycle-dependent closed ICME magnetic-flux component.\\  
\tab \hspace{5mm}Alternatively, total open flux can be determined by identifying coronal holes in chromospheric or coronal observations.  Coronal hole areas are calculated from observations after their boundaries are determined either through manual \citep{HarveyRecely2002} or automated ({\it e.g.} \citealp{HenneyHarvey2005,SchollHabbal2008,KristaGallagher2009,Lowder2014,Verbeeck2014,Boucheron2016,Caplan2016,Garton2018,Hamada2018}) methods.  To derive open flux, coronal-hole contours are paired with corresponding synoptic photospheric-field maps. The advent of high- resolution coronal EUV images allows for an independent estimate of the open flux in comparison to model estimates and {\it in-situ} observations for the past decade. Observationally derived coronal holes also provide constraints on model-derived coronal hole boundaries.\\
\tab \hspace{5mm}In this article, we estimate total open magnetic flux from 1990\,--\,2013 in two different ways using global photospheric magnetic-field maps derived from National Solar Observatory (NSO) Kitt Peak Vacuum Telescope (KPVT: \citealp{KPVT-Jones1992}) and Vector Spectromagnetograph (VSM: \citealp{VSM-Henney2009}) observations.  While data prior to 1990 are available from KPVT, their calibration before this time is suspect \citep{Arge2002} and therefore they are not used in this study.  First, traditional Carrington (or diachronic) NSO/KPVT and VSM maps are used as input for the Wang--Sheeley--Arge (WSA) model \citep{ArgePizzo2000,Arge2003a,Arge2004}, to estimate total open heliospheric flux, as are Air Force Data Assimilative Photospheric Flux Transport (ADAPT:  \citealp{ADAPT_Arge2009,Arge2010,Arge2013,Hickmann2015}) synchronic maps generated from the same NSO/KPVT and VSM magnetograms. Second, the same photospheric-field maps are paired with EUV and He {\sc I} 10830 \AA\ manually derived coronal holes to estimate total open flux and provide an objective comparison.  The results of these two methods are then compared to the open heliospheric flux derived from daily averaged $|B_\text{r}|$ spacecraft measurements.  Daily averages were used before taking the unsigned value in lieu of a more advanced technique to correct for kinematic effects, because both methods have been shown to produce comparable open-flux estimates (\citealp{Lockwood2009c} Figure 9).  We address the following questions: \begin{enumerate}[i)]
  \item Does WSA produce significantly different open-flux estimates when using diachronic \textit{vs.} synchronic photospheric-field maps derived from identical magnetograms from the same observatories?
  \item How do open-flux estimates based on observationally derived coronal holes compare with those derived with a) WSA {\it when using the same photospheric-field maps} and b) single-point measurements from spacecraft?
   \item What can open flux obtained from coronal-hole observations reveal in regards to potential sources of disagreement between model-derived and spacecraft-observed estimates?
\end{enumerate}
\tab \hspace{2mm}This work is thus an independent check on results obtained using WSO and MWO input maps with additional correction factors applied \citep[see][]{Wang2000,Riley2007}.  Similar to our work, \citet{Linker2016} used a PFSS model driven by ADAPT-VSM maps to estimate $B_\text{r}$ at 1 AU and obtained good agreement between model-derived and spacecraft-observed $B_\text{r}$ if the ADAPT-VSM maps were scaled by 1.5.  In our study we do not apply any correction factors to the photospheric-field maps. Complementary studies involving the use of EUV observations to determine total heliospheric flux have been conducted by \citet{Linker2017} and \citet{Lowder2017}. In these cases, open-flux estimates obtained from observationally derived coronal holes were compared with that derived from {\it in-situ} observations and models, but coronal holes were identified through automated detection methods.  Here, we manually identify coronal holes in EUV to calculate total open flux. We also compare open-flux estimates calculated by \citet{HarveyRecely2002} from manually derived coronal holes using He {\sc I} 10830 \AA\ observations. \\
\tab \hspace{5mm} This article is organized as follows: Section~\ref{2-ModelsVSobs} describes the specific models and photospheric-field maps used in this study and then compares the total open flux derived from conventional methods. Section~\ref{3-CHobs} introduces the He and EUV observationally derived coronal hole method for obtaining open flux.  The results from all methods are compared and discussed in Section~\ref{4-Results}, and summarized in the final section. 

%%%%%%%%%%%%%%%%%%%%%%%%%%%%%%%%%%%%%%%%%%%%%%%%%%%%%%%%%%%%%%%%%
%%%%%%%%%%%%%%%%%%%     Begin Section 2   %%%%%%%%%%%%%%%%%%%%%%%
%%%%%%%%   Models vs. {\it in situ} Obs for Calc. Open Flux  %%%%%%%%%%
%%%%%%%%%%%%%%%%%%%%%%%%%%%%%%%%%%%%%%%%%%%%%%%%%%%%%%%%%%%%%%%%%

\section{Model Estimation of Open Flux}
\label{2-ModelsVSobs}

\begin{figure}  
  \centerline{\includegraphics[width=1\textwidth,trim={.4cm .7cm .4cm 5cm},clip]{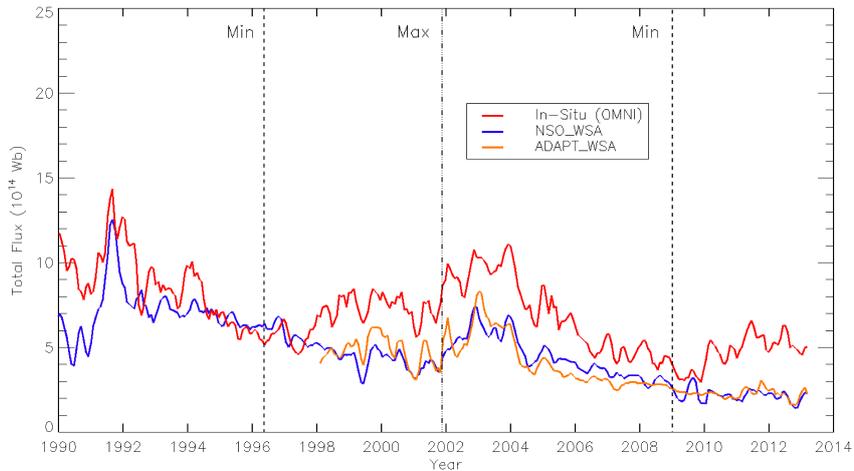}} %%% To Trim: <left> <lower> <right> <upper>
   %\centerline{\includegraphics[width=1.08\columnwidth,trim={0cm .9cm 0.5cm 24.0cm},clip]{open_flux_Figure_1.eps}} %%% To Trim: <left> <lower> <right> <upper>%%% To Trim: <left> <lower> <right> <upper>
   \caption{Total open flux, smoothed with running mean with a temporal window of three rotations, derived from WSA using NSO diachronic (\textit{blue line}) and synchronic ADAPT (\textit{orange line}) maps, along with {\it in-situ} observations (\textit{red line}).}
   \label{Figure1}
   \end{figure}

In studies of the temporal variation of open magnetic flux, it is common to consider either the flux of one polarity or the total of the unsigned flux ({\it i.e.} using the absolute value of the radial field) in order to have nonzero values. The total unsigned open heliospheric flux is used here, and every mention of “total open flux” throughout the rest of this article to the total unsigned flux value. This section describes the WSA model used to estimate total open flux, discusses the two types of photospheric magnetic-field maps ({\it i.e.} diachronic \textit{vs.} synchronic) used as model input, and then compares the results obtained with WSA to those derived from spacecraft observations. \\  
\tab \hspace{5mm}Using global photospheric magnetic-field maps as input, the WSA model derives a standard PFSS solution of the coronal field from the photosphere out to the source-surface height of 2.5 solar radii (\(\text{R}_\odot\)). Potential field models are simple, efficient, and known to give similar open magnetic-flux estimates to that obtained from more complex MHD solutions \citep{deToma2005,MHDvsPFSS-Riley2006}. PFSS model estimates have been shown to agree overall between near-Earth observed and PFSS-derived IMF \citep{SSHeight-Hoeksema1983}; however, recent studies suggest that varying the source-surface height throughout a solar cycle may be more appropriate \citep{Lee2011,Arden2014}. \\  
\tab \hspace{5mm}The data input for WSA, and other coronal models, are estimates of the global photospheric magnetic-field distribution. Traditional global magnetic maps, sometimes referred to as Carrington or diachronic synoptic maps, are comprised of remapped full-disk magnetogram observations accumulated over a 27.2753 day synodic (or Carrington) rotation period. Diachronic maps are prepared in a variety of ways, such as taking a narrow longitudinal width slice along central meridian for each magnetogram, remapping the slice into heliographic coordinates, and then assembling them in sequence over the period of a solar rotation.  This process results in an unrealistic mixture of space and time, since known magnetic-field transport processes ({\it e.g.} differential rotation, supergranulation flows, meridional drifts, \textit{etc.}) are not applied \citep[\textit{e.g.}][]{WordenHarvey2000}. Thus, traditional synoptic maps are diachronic in nature because they represent a history of the central meridian over a Carrington rotation and not the photospheric magnetic-field distribution at a fixed moment in time \citep{Linker2013}. \\ 
\tab \hspace{5mm}Flux-transport models ({\it e.g.} \citealp{WordenHarvey2000,SchrijverDeRosa2003}), such as ADAPT, are more realistic alternatives to traditional synoptic maps, since they provide a synchronic (instantaneous) representation of the global photospheric magnetic field. ADAPT is a modified version of the \citet{WordenHarvey2000} flux-transport model, which also incorporates various data-assimilation methods, enabling flux to be evolved where there is a lack of observations.  The photospheric field is then updated through assimilating new observations as they become available.  It takes into account differential rotation, poleward meridional flows, supergranulation, and small-scale random flux emergence. The ADAPT model data assimilation accounts for the model ensemble and observational data uncertainties to generate global magnetic maps \citep{Hickmann2015}. For instance, ADAPT heavily weights observations taken near the central meridian where observations are most reliable, while the model specification of the field is generally given more weight near the limbs where observations are the least reliable. ADAPT is an ensemble model producing several ({\it i.e.} 12 for this study) realizations of the global magnetic field representing, as realistically as possible, the range of possible states of the global solar photospheric magnetic-flux distribution at any given moment in time. When coupled with WSA, ADAPT-WSA derives an ensemble of solutions for the coronal field and solar wind speed at 1 AU.  This model output can then be compared to observations to determine the most accurate coronal solution of the 12 realizations. The ensemble of photospheric-field solutions produced by ADAPT can thus be used to determine the most realistic photospheric boundary conditions for a certain time period.\\
\tab \hspace{5mm}The total open flux from 1990 to 2013, shown in Figure \ref{Figure1}, is determined from daily averaged $|B_\text{r}|$ spacecraft measurements near 1 AU (\url{omniweb.gsfc.nasa.gov/}), and from WSA using both synchronic ADAPT maps (\url{ftp://gong2.nso.edu/adapt/maps}) and diachronic Carrington KPVT and VSM maps (\url{solis.nso.edu/0/vsm/vsm_maps.php}) as its input. The ADAPT-WSA results begin in 1998 due to known calibration issues with KPVT magnetograms prior to this year. Calibration offsets between different magnetogram sources are known to range widely and can present a significant challenge in studies like this (\citealp{Riley2014}, Table 3). To avoid any issues with potential magnetogram offsets, the photospheric-field maps used with WSA to estimate total open flux were created using the same set of VSM (September 2003 onward) and KPVT (August 2003 and prior) magnetograms, yet the two types of maps ({\it i.e.} synchronic and diachronic) were generated from entirely different approaches. \\
\tab \hspace{5mm}All of the data sets are plotted as three-Carrington-rotation running averages. For the spacecraft observations, this meant averaging $\approx$80 daily values ({\it i.e.} three rotation's worth) of the unsigned radial magnetic field and then calculating the total open flux.  When using WSA with diachronic maps, the total open flux is derived for each rotation and then averaged over three Carrington rotations.  When using ADAPT maps with WSA, the total unsigned open flux was calculated for the mid-rotation date 
%%%%%%%%%%%%%%%%%%%%%%%%%%%%%%%%%%%%%%%%%%%%%%%%%%%%%%%%%%%%%%%
%%%%%%%%%%%%%%%%%%%%  Insert Figure 1  %%%%%%%%%%%%%%%%%%%%%%%%
%%%%%%%%%%%%%%%%%%%%%%%%%%%%%%%%%%%%%%%%%%%%%%%%%%%%%%%%%%%%%%%
%%%%%%%%%%%%%%%%%%%%%%%%%%%%%%%%%%%%%%%%%%%%%%%%%%%%%%%%%%%%%%%
%%%%%%%%%%%%%%%%%%%%  Insert Figure 1  %%%%%%%%%%%%%%%%%%%%%%%%
%%%%%%%%%%%%%%%%%%%%%%%%%%%%%%%%%%%%%%%%%%%%%%%%%%%%%%%%%%%%%%%
of each Carrington rotation.  For any given moment in time, ADAPT produces 12 realizations for the photospheric field that represent the uncertainty in time-dependent phenomena.  Therefore, over a three-Carrington-rotation running average the results using ADAPT maps as input into WSA include 36 calculations of the total open flux ({\it i.e.} 12 for each rotation).\\
\tab \hspace{5mm}The WSA model-derived open flux using both synchronic and diachronic maps track each other reasonably well for the entire fifteen-year period of available data. However, on average ADAPT-WSA yields fluxes that are slightly higher than those from traditional NSO diachronic maps using KPVT magnetograms, while the reverse appears to be largely true for the VSM magnetograms, at least until 2009 and beyond when the two agree rather well. In retrospect this result is not entirely surprising, as the synchronic and diachronic maps are generated using the same input magnetograms and the results are averaged over three Carrington rotations. So while day-to-day variations are very likely, it is not surprising that the two sets of maps generally agree on three-rotation or $\approx$80 day timescales.\\ 	
\tab \hspace{5mm}While total open magnetic flux estimates determined using {\it in-situ} observations agree somewhat with those derived using WSA from 1992 to 1998 (see Figure \ref{Figure1}), from 1998 onward they are consistently higher than the model-derived results.  This deviation is most pronounced during solar maximum. It is difficult to investigate the source(s) of these discrepancies without an alternative and independent method to estimate open flux.  Such alternative and objective methods can also be used to provide constraints to model parameters.  The following section discusses an independent approach for deriving open flux using readily available global observations of the corona. 
%%%%%%%%%%%%%%%%%%%%%%%%%%%%%%%%%%%%%%%%%%%%%%%%%%%%%%%%%%%%%%%%%
%%%%%%%%%%%%%%%%%%%     Begin Section 3   %%%%%%%%%%%%%%%%%%%%%%%
%%%%%%%%%%%    Estimating Open Flux Using CH Obs  %%%%%%%%%%%%%%%
%%%%%%%%%%%%%%%%%%%%%%%%%%%%%%%%%%%%%%%%%%%%%%%%%%%%%%%%%%%%%%%%%

\section{Estimating Open Flux Using coronal-hole observations} 
      \label{3-CHobs}      

Global diachronic maps have been constructed from He {\sc I} 10830 \AA\ \citep{HarveySheeley1977} and EUV \citep{EIT-Delaboudiniere1995} observations, beginning in the late 1970's. More recently, combining observations from the \textit{Solar TERrestrial RElations Observatory} (STEREO: \citealp{STEREO-Kaiser2008}) and the \textit{Solar Dynamics Observatory} (SDO: \citealp{SDO-Pesnell2012}) allows for the generation of global synchronic maps in EUV.  Both diachronic and synchronic maps of the global corona can be used to identify coronal holes and, assuming the majority of open magnetic field resides within coronal hole boundaries, obtain the total open magnetic flux for an entire rotation or day respectively.  This is accomplished through calculating the flux within each pixel inside a coronal hole by multiplying the radial-magnetic-field value with the differential area on the solar surface (assumed to be spherical) at that location, and then summing individual flux values over the entire map, using the following equation: 
\begin{eqnarray}  
 \label{Open_Flux_defn}
 \Phi = \sum_{i,j}^n c_{i,j} |\bb_{i,j}|r^2\sin(\theta)\Delta\theta\Delta\phi
   \end{eqnarray}
where $|\bb_{i,j}|$ is the unsigned-radial-field component at a given point on the photosphere ({\it i.e.} pixel of a global coronal image), $r^2\sin(\theta)\Delta\theta\Delta\phi$ is the differential surface area [\textit{dA}], and $c_{i,j}$ equals one or zero, depending on whether the given point is inside (one) or outside (zero) of a coronal hole.\\ 
\tab \hspace{5mm}In this study, coronal holes were identified observationally on global diachronic He {\sc I} 10830 \AA\ and synchronic EUV coronal maps and paired with the corresponding type of photospheric-field maps to calculate the total open flux using Equation \ref{Open_Flux_defn}. The two sets of photospheric-field maps were the same ones used in WSA to calculate the total open flux.  Although this approach is not altogether new ({\it e.g.} \citealp{HarveyRecely2002}), it provides an alternative method of obtaining the total open flux and thus can be used as an independent check to compare with both model-derived and spacecraft-observed open flux.  

\subsection{Helium \textsc{I} 10830 \AA\ Observations} %%%%%%%%%%%%%%
  \label{3.1-He1083_Obs}

The technique of observationally identifying coronal holes using the equivalent width of the He {\sc I} 10830 \AA\ line has been used to study polar coronal-hole evolution during Solar Cycles 22 and 23.  For more than a decade \citet{HarveyRecely2002} manually identified coronal holes from NSO/KPVT He {\sc I} 10830 \AA\ spectroheliograms created from full-disk observations.  He {\sc I} 10830 \AA\ is a chromospheric spectral line in which the coronal radiation overpopulates the multiplet transition levels.  This process increases scattering and absorption in areas where coronal emission is high.  In coronal holes (where coronal emission is low), the opposite effect occurs and chromospheric He {\sc I} absorption decreases (\citealp{HarveyRecely2002}, and references therein).  Coronal holes are thus visible in He {\sc I} 10830 \AA\ as bright regions relative to the background emission, as opposed to dark in EUV/X-ray.  Harvey and Recely identified these boundaries of low and high chromospheric contrast, and they manually contoured coronal holes that were between 75\,--\,100\,\% unipolar and at least two supergranules (10$^9$ km$^2$) in size.  They used their contours to classify coronal holes as either isolated ({\it i.e.} mid-latitude) or polar, and determined such properties as area and magnetic flux.  They calculated flux within coronal holes by overlaying their contoured coronal-hole maps onto KPVT global maps 
%%%%%%%%%%%%%%%%%%%%%%%%%%%%%%%%%%%%%%%%%%%%%%%%%%%%%%%%%%%%%%%
%%%%%%%%%%%%%%%%%%%%  Insert Figure 2  %%%%%%%%%%%%%%%%%%%%%%%%
%%%%%%%%%%%%%%%%%%%%%%%%%%%%%%%%%%%%%%%%%%%%%%%%%%%%%%%%%%%%%%%
  \begin{figure}   
  \hspace*{-.5cm}
   \centerline{\includegraphics[width=1\textwidth,trim={1.5cm 2.8cm .7cm 3.3cm},clip]{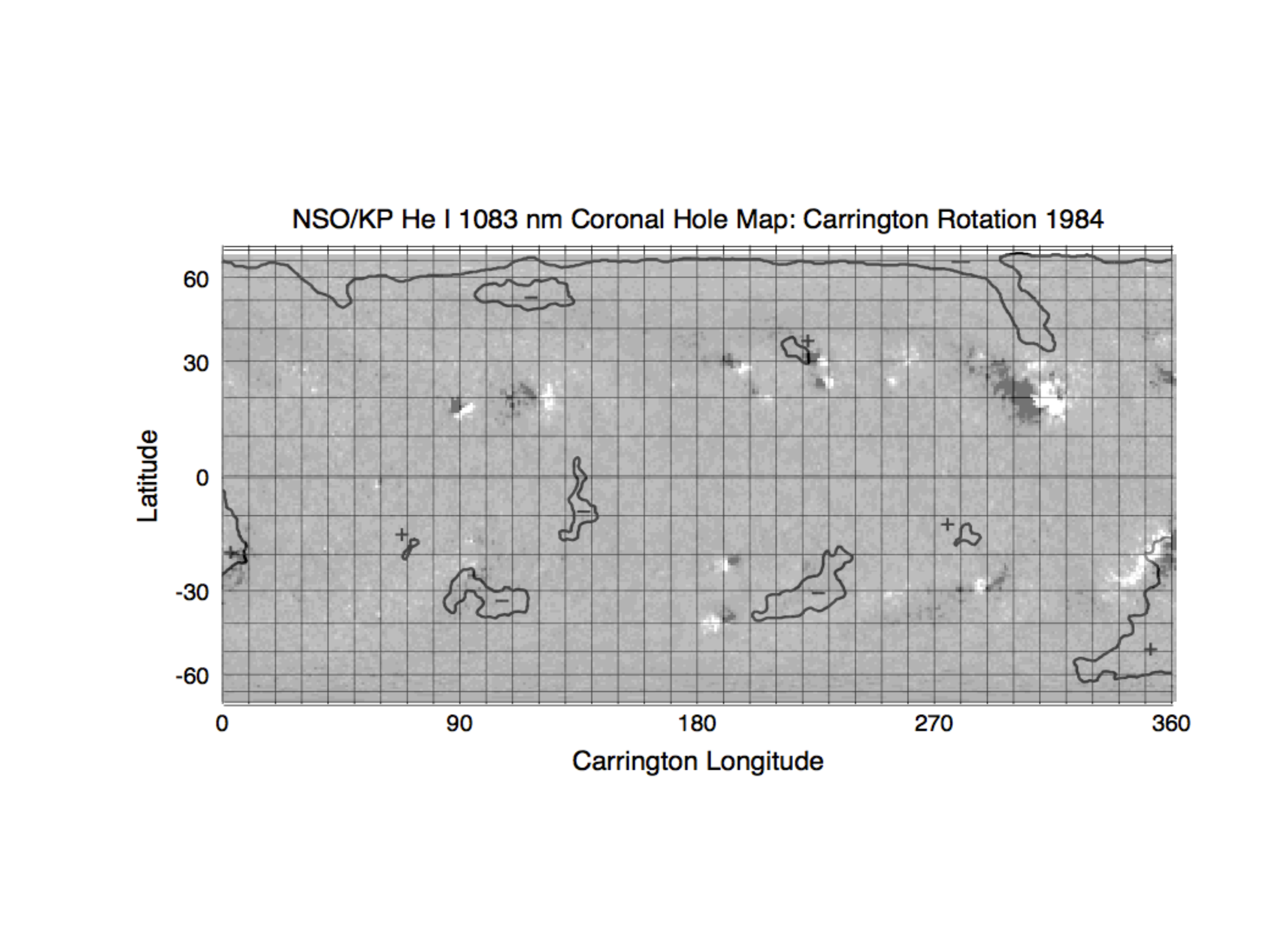}} %%% To Trim: <left> <lower> <right> <upper>
              \caption{Diachronic coronal-hole maps (\textit{contours}) created from He {\sc I} 10830 \AA\ spectroheliograms \citep{HarveyRecely2002} overlaid on global KPVT photospheric-field (\textit{grayscale}) maps to calculate total open flux as defined in Equation \ref{Open_Flux_defn}.}
  		 \label{Figure2}
   \end{figure}
%%%%%%%%%%%%%%%%%%%%%%%%%%%%%%%%%%%%%%%%%%%%%%%%%%%%%%%%%%%%%%%
%%%%%%%%%%%%%%%%%%%%  Insert Figure 2  %%%%%%%%%%%%%%%%%%%%%%%%
%%%%%%%%%%%%%%%%%%%%%%%%%%%%%%%%%%%%%%%%%%%%%%%%%%%%%%%%%%%%%%%
of the photospheric field (Figure \ref{Figure2}).  Together the NSO/KPVT spectroheliograms and global magnetic maps make a suitable pairing because both are diachronic and observed with the same telescope. Harvey and Recely's original coronal-hole and open magnetic-flux calculations (available at \url{ftp://solis.nso.edu/kpvt/coronal_holes/synoptic/}) span from 1989 to 2002, and were directly used in this study without modification.\\  
\tab \hspace{5mm}Figure \ref{AllResults-Plot} displays the results previously shown for total open heliospheric flux derived from i) {\it in-situ} observations, ii) the WSA model using two different input maps (diachronic and ADAPT synchronic), and now including iii) the Harvey and Recely derived open fluxes (calculated with diachronic photospheric-field maps) from 1989 to 2003, and iv) results obtained by applying Harvey and Recely's technique to global synchronic EUV coronal maps (discussed in Section \ref{3.2-EUV_Obs}). Figure \ref{AllResults-Plot} highlights excellent agreement between the open flux calculated from He 10830 \AA\ observations and the WSA values for the nine-year period (1992\,--\,2000) centered roughly around solar minimum.  This suggests that the model captures well the coronal holes identified observationally, at least over three-Carrington-rotation timescales. A portion of this time period (approximately 1994\,--\,1998) is also when the spacecraft data agree best with the results from the other methods.  Near solar maximum (1990/2002) the flux estimates from {\it in-situ} observations are consistently higher than that obtained from both the WSA model and the derived coronal hole method.\\  
\tab \hspace{5mm}Smaller discrepancies ({\it i.e.} compared to the {\it in-situ} results) are exhibited in Figure \ref{AllResults-Plot} between the WSA model-derived open flux and that deduced from helium observations near both periods of solar maximum (approximately 1990\,--\,1992 and late 1999\,--\,2002).  These differences could be due to a combination of factors, such as the known difficulty of using He observations to observe coronal holes at the solar mid-latitude region during periods of high activity \citep{Arge2003b}.  Near solar maximum, mid-latitude coronal holes can have field strengths up to four times stronger
compared to those observed during minimum \citep{Arge2002}.  If some mid-latitude coronal holes were undetected during solar maximum, it could explain why the He-derived open flux is lower than WSA model predicted values for these years.
%%%%%%%%%%%%%%%%%%%%%%%%%%%%%%%%%%%%%%%%%%%%%%%%%%%%%%%%%%%%%%%
%%%%%%%%%%%%%%%%%%%%  Insert Figure 3  %%%%%%%%%%%%%%%%%%%%%%%%
%%%%%%%%%%%%%%%%%%%%%%%%%%%%%%%%%%%%%%%%%%%%%%%%%%%%%%%%%%%%%%%
  \begin{figure}  
  %\hspace{-.1cm}
  \centerline{\includegraphics[width=1\textwidth,trim={.4cm .7cm .4cm 5cm},clip]{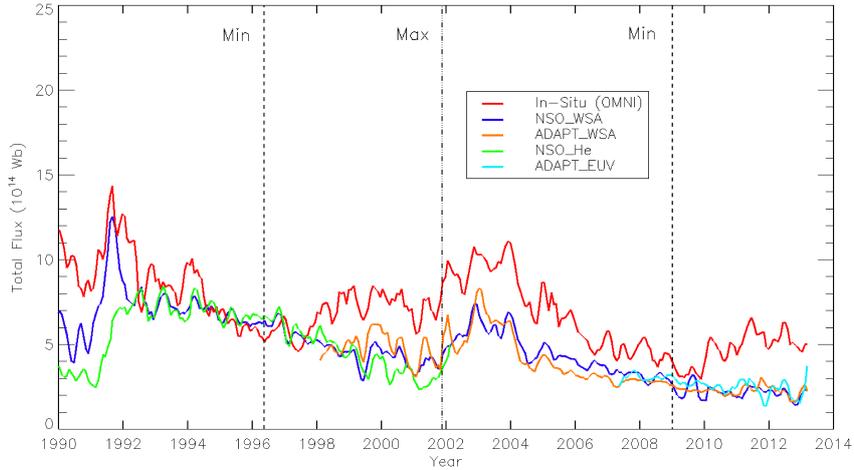}} %%% To Trim: <left> <lower> <right> <upper>
   %\centerline{\includegraphics[width=1.08\columnwidth,trim={0cm .9cm 0.5cm 24.0cm},clip]{open_flux_Figure_3.eps}} %%% To Trim: <left> <lower> <right> <upper>
              \caption{Total open flux, smoothed with running mean with a temporal window of three rotations, derived from WSA using NSO diachronic (\textit{blue line}) and synchronic ADAPT (\textit{orange line}) maps, {\it in-situ} observations (\textit{red line}), and observations of He and EUV coronal holes (\textit{green} and \textit{cyan} lines respectively). }
              \label{AllResults-Plot}
   \end{figure}
   
 %  As in Figure 1, with the addition of total open flux derived from observations of He and EUV coronal holes (\textit{green} and \textit{cyan} lines respectively). }

%%%%%%%%%%%%%%%%%%%%%%%%%%%%%%%%%%%%%%%%%%%%%%%%%%%%%%%%%%%%%%%
%%%%%%%%%%%%%%%%%%%%  Insert Figure 3  %%%%%%%%%%%%%%%%%%%%%%%%
%%%%%%%%%%%%%%%%%%%%%%%%%%%%%%%%%%%%%%%%%%%%%%%%%%%%%%%%%%%%%%%
Overall, the Harvey and Recely approach provides a method of deriving the total open flux that is independent of coronal models and, nonetheless, is found to match WSA results. In addition to diachronic He {\sc I} 10830 \AA\ spectroheliograms, we use EUV synchronic maps (from STEREO and SDO) to manually identify coronal holes, described in the following section, and calculate total open flux by applying Harvey and Recely's technique. The coronal-hole identification process is also explained in the next section, as well as how open flux was calculated by pairing EUV-derived coronal holes with ADAPT photospheric-field maps.\\

\subsection{EUV Observations} %%%%%%%%%%%%%%
\begin{figure} 
  %\hspace*{.01cm}
   %\centerline{\includegraphics[trim={0cm 2cm 0.2cm 2cm},clip]
   \centerline{\includegraphics[width=1.1\textwidth,trim = {0cm 2.3cm .5cm 3.7cm},clip]
   {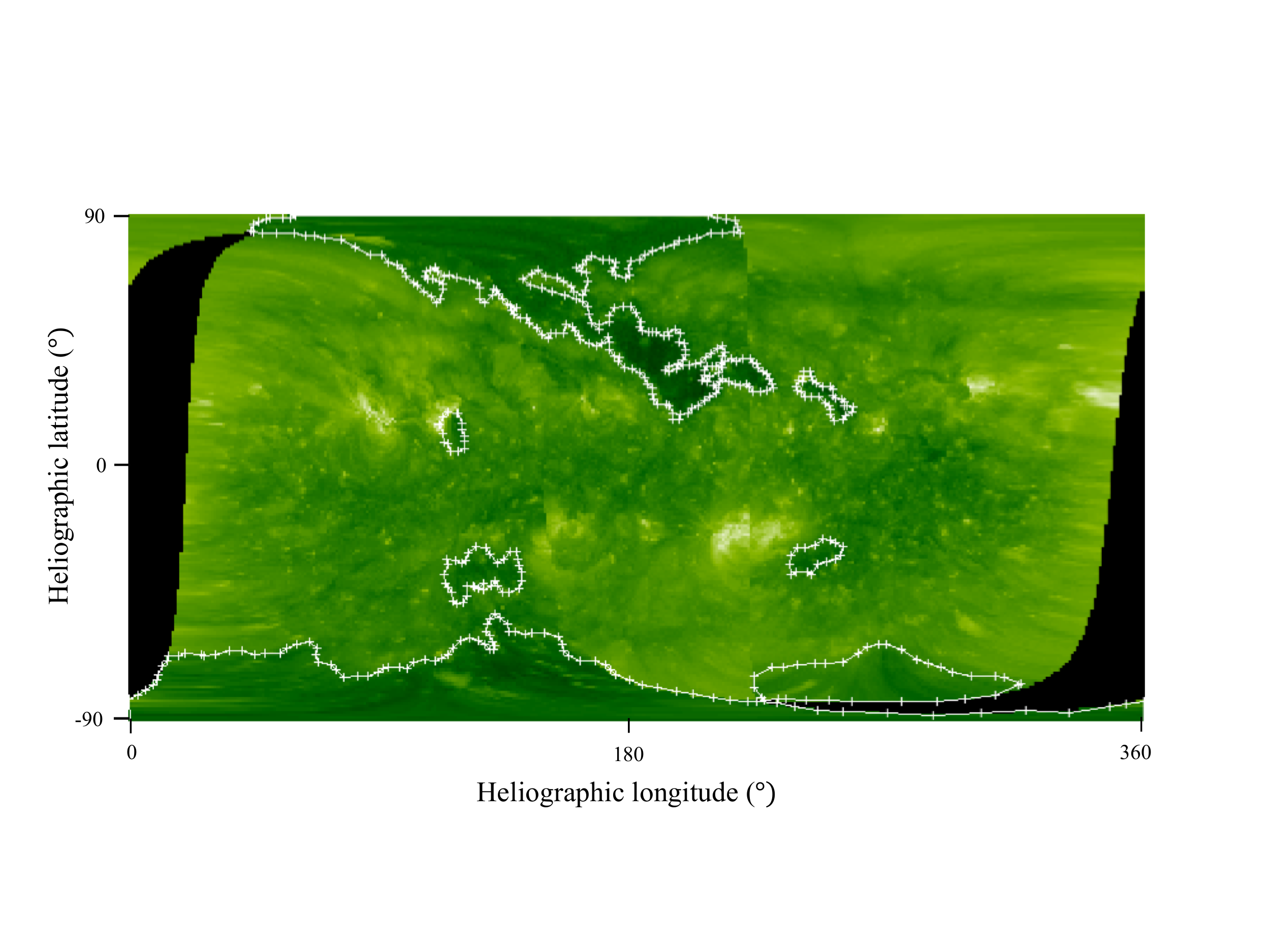}} %%% To Trim: <left> <lower> <right> <upper>   
              \caption{Global heliographic synchronic EUV map of the corona with contoured coronal holes for 27 July 2010 (CR 2099 mid-rotation date).  Maps were assembled using disk images from STEREO/EUVI B (left), SDO/AIA (middle), and STEREO/EUVI A (right).}
               \label{EUVmap} 
   \end{figure}

  \label{3.2-EUV_Obs}
The STEREO spacecraft consists of two identical observatories launched in October 2006, one trailing behind Earth's orbit (STEREO-B) and the other ahead (STEREO-A).  The \textit{Sun Earth Connection Coronal and Heliospheric Investigation} (SECCHI) \textit{Extreme Ultraviolet Imager} (EUVI: \citealp{SECCHI-Howard2008}) observes the Sun from the chromosphere to the inner corona ($\leq$1.7 \(\text{R}_\odot\)) via four EUV emission lines: 171, 195, 284, and 304 \AA.  In February 2011, STEREO-A and -B were 180\degree\ apart with each about 90\degree\ from the Sun--Earth line. When combined, observations from the two spacecraft provided the first global synchronic representation of the solar corona. SDO was launched 
in February 2010 in geosynchronous orbit. One of the instrumentation suites, the \textit{Atmospheric Imaging Assembly} (AIA: \citealp{AIA-Lemen2012}) consists of four telescopes, of which there are seven EUV channels ranging from 94 to 335 \AA\ ($6 \times 10^4$ to $2 \times 10^7$ K).\\
\tab \hspace{5mm}Synchronic EUV maps (Figure \ref{EUVmap}) corresponding to the first and mid-rotation dates of CRs 2056 to 2135 (March 2007\,--\,April 2013) were generated using the {\sf{SolarSoftware}} \citep{SolarSoft} routine {\sf{ssc\_euvi\_synoptic.pro}}.  The Earth--Sun central meridian longitude ({\it i.e.} \ang{0} or \ang{180}) corresponding to each of these dates was positioned at the center of each map.  STEREO/EUVI Fe {\sc XII} 195 \AA\ full-disk observations from the A and B spacecraft were used to create each map. As the  twin spacecraft separated in angle from 2007 to 2011, each map became increasingly more global in coverage.  However, the increased spacecraft separation also resulted in a larger portion of each EUV map being subject to smoothing and line-of-sight effects ({\it e.g.} at the limbs) from remapping full-disk observations. To mitigate this problem significantly and provide more complete coverage, when available (March 2010 onward) SDO/AIA Fe {\sc XII} 193 \AA\ observations were used to replace the STEREO observations at the solar limb with SDO/AIA disk-center observations.\\ 
\tab \hspace{5mm}In order to derive total open flux from a synchronic EUV map, i) coronal holes have to be identified, and ii) the magnetic field at each pixel inside a coronal hole must be summed over the entire map using Equation \ref{Open_Flux_defn}.  For this process to be effective, global EUV coverage is necessary. Due to the varying angular separation of STEREO-A and -B this is not always possible, especially from 2007\,--\,2010.  
Each synchronic map constructed from STEREO/EUVI observations always has complete coverage (except for small portions at the poles) within the $\pm$ 90 degrees surrounding central meridian, but the outer \ang{90} of any map is subject to observational gaps (black regions in Figure \ref{EUVmap}) that must be accounted for to determine total open flux.  As explained in detail below, this is resolved 
%%%%%%%%%%%%%%%%%%%%%%%%%%%%%%%%%%%%%%%%%%%%%%%%%%%%%%%%%%%%%%%
%%%%%%%%%%%%%%%%%%%%  Insert Figure 4  %%%%%%%%%%%%%%%%%%%%%%%%
%%%%%%%%%%%%%%%%%%%%%%%%%%%%%%%%%%%%%%%%%%%%%%%%%%%%%%%%%%%%%%%
%%%%%%%%%%%%%%%%%%%%%%%%%%%%%%%%%%%%%%%%%%%%%%%%%%%%%%%%%%%%%%%
%%%%%%%%%%%%%%%%%%%%  Insert Figure 4  %%%%%%%%%%%%%%%%%%%%%%%%
%%%%%%%%%%%%%%%%%%%%%%%%%%%%%%%%%%%%%%%%%%%%%%%%%%%%%%%%%%%%%%%
by first contouring coronal holes on each EUV synchronic map and then calculating total open flux for each CR from a composite map of the first, middle, and last day of the rotation.\\
\tab \hspace{5mm}All of the constructed synchronic maps ({\it e.g.} Figure \ref{EUVmap}), one corresponding to the first and another the mid-rotation date of each CR, were used to manually identify coronal holes, similarly to the approach of Harvey and Recely.   Each synchronic map was first inspected for concentrated areas of darkness.  At EUV wavelengths, coronal holes are areas of reduced emission and appear visually as dark regions against the background corona. At the solar limb, coronal holes appear brighter in EUV emission than those at disk center due to obscuration from extended emission and smoothing, and they are more difficult to identify. Individual full-disk observations from STEREO/EUVI and SDO/AIA were used to examine the limbs more closely to determine if coronal holes were present. This approach of calculating open flux assumes that magnetically open  regions on the Sun are dark in EUV emission.\\  
\tab \hspace{5mm}A computer software \citep{Pattichis2014} was used to manually draw contours around each coronal hole.  This software also provided the ability to overlay magnetic neutral lines on each EUV map to distinguish between coronal holes ({\it i.e.} open field regions) and filaments ({\it i.e.} closed field regions), because both have reduced EUV emission and are visually dark.  For coronal holes identified in the AIA portion of a synchronic map, three-color channel SDO/AIA composite images provided by the Lockheed Martin Solar and Astrophysics Laboratory (LMSAL) were an additional tool used to rule out filaments.  These full-disk images are an overlay of Fe {\sc XIV} 211 \AA, Fe {\sc XII} 193 \AA, and the Fe {\sc IX} 171 \AA\ lines in three different hues, red, green, and blue respectively.  Coronal holes in these images appear deep blue because they have 
 %%%%%%%%%%%%%%%%%%%%%%%%%%%%%%%%%%%%%%%%%%%%%%%%%%%%%%%%%%%%%%%
%%%%%%%%%%%%%%%%%%%%  Insert Figure 5  %%%%%%%%%%%%%%%%%%%%%%%%
%%%%%%%%%%%%%%%%%%%%%%%%%%%%%%%%%%%%%%%%%%%%%%%%%%%%%%%%%%%%%%%
  \begin{figure}   
  %\hspace*{.01cm}
   %\centerline{\includegraphics[trim={0cm 2cm 0.2cm 2cm},clip]
   \centerline{\includegraphics[width=.98\textwidth, trim = {.9cm .5cm 1.2cm 2.0cm},clip]
   {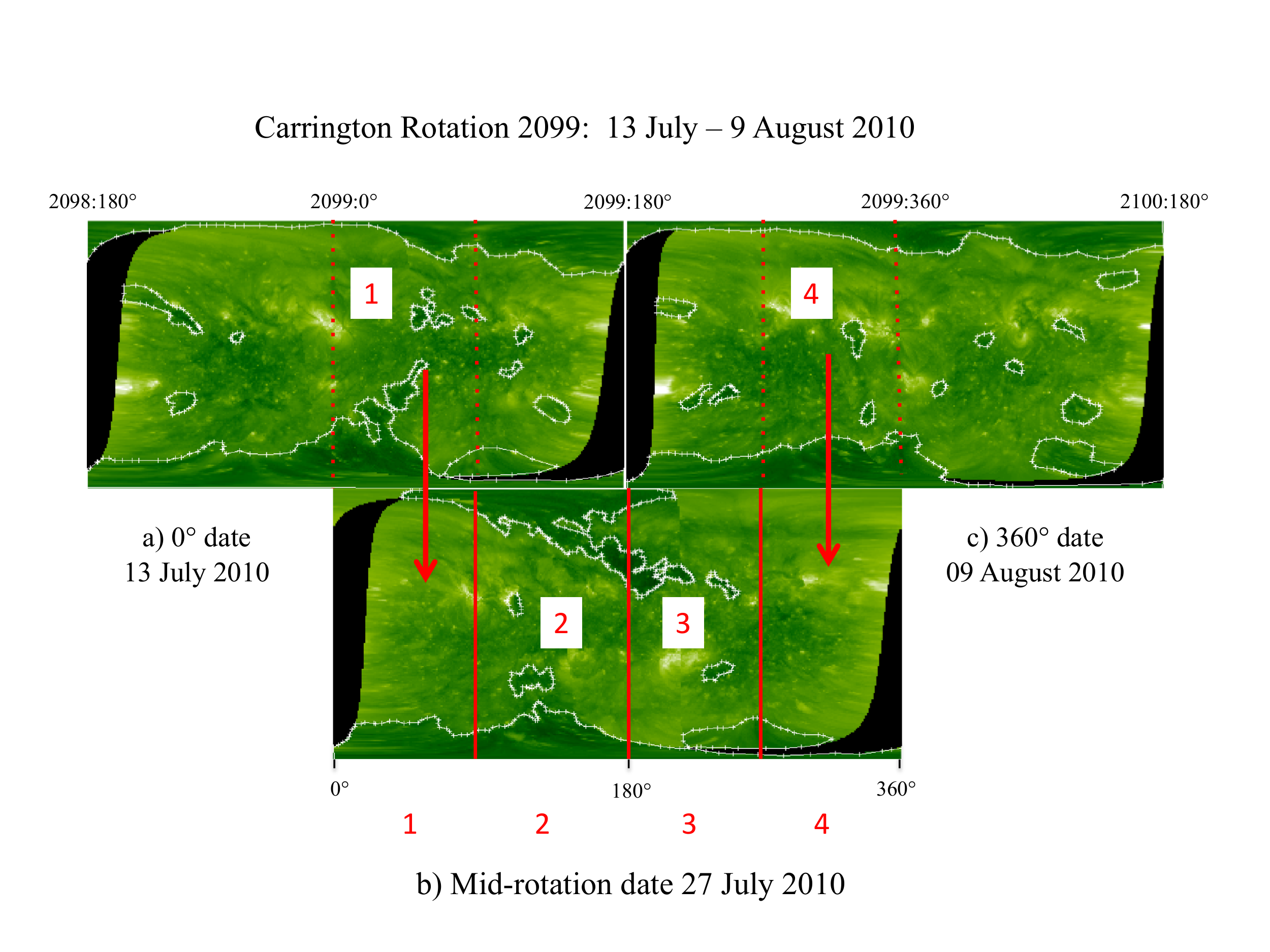}} %%% To Trim: <left> <lower> <right> <upper> 
              \caption{Observationally derived coronal holes on original EUV synchronic maps corresponding to a) first, b) middle, and c) last ({\it i.e.} first day of next) day of CR 2099. This rotation is used as an example to illustrate how EUV observations were used to calculate total open flux for the midpoint of each CR despite periods with insufficient observations, as discussed in the text.}
        \label{EUVcutandstitch}
   \end{figure}
%%%%%%%%%%%%%%%%%%%%%%%%%%%%%%%%%%%%%%%%%%%%%%%%%%%%%%%%%%%%%%%
%%%%%%%%%%%%%%%%%%%%  Insert Figure 5  %%%%%%%%%%%%%%%%%%%%%%%%
%%%%%%%%%%%%%%%%%%%%%%%%%%%%%%%%%%%%%%%%%%%%%%%%%%%%%%%%%%%%%%%
such low emission, but emit slightly more at 171 \AA\ \citep{Garton2018}.  In contrast, filaments have more of a reddish hue in the LMSAL composite images because the predominant emission is at 211 \AA\ \citep{AIA-Lemen2012}.\\   
\tab \hspace{5mm}Visually dark regions were manually contoured once they were confirmed to be coronal holes.  The STEREO/EUVI and SDO/AIA original full-disk observations were used as an aid to provide a more realistic depiction of the true shape of the identified coronal holes.  An example of a contoured map is shown in Figure 4. The Pattichis software generated binary files with \ang{2} resolution for each contoured EUV map that consisted of a one (zero) for pixels inside (outside) a coronal hole.  Each binary file was paired with the corresponding ADAPT map in order to calculate total open flux using Equation \ref{Open_Flux_defn}. This pairing is different than Harvey and Recely's original study in that both the contoured coronal holes and the photospheric field are synchronic.\\ 
\tab \hspace{5mm}As mentioned previously, the original synchronic maps ({\it e.g.} Figure \ref{EUVmap}) did not always provide global EUV coverage due to the angular separation of STEREO-A, -B, and SDO. However, observations from the three spacecraft at a minimum provide full coverage within the central \ang{180} portion of each map.  Using \ang{90} segments from this region of contoured EUV maps representing the first, middle, and last ({\it i.e.} first date of the next CR) date of each CR, a composite map is created representing each rotation's midpoint. 
Composite maps are then used to calculate total open flux for each Carrington rotation.  Figure \ref{EUVcutandstitch} uses CR 2099 to illustrate this process.\\
\tab \hspace{5mm}Figure 5a and b are the contoured coronal-hole maps for the commencement date and mid-rotation date of CR 2099 respectively.  Figure 5c is the map for the first day of next rotation, which was used to represent the last day of CR 2099.  The total open flux was calculated for each of these dates individually using the original maps.  Panel 1 in Figure 5a represents the beginning of CR 2099 from \ang{0} to \ang{90}.  Panels 2 and 3 for the mid-rotation date (Figure 5b) represent CR 2099 from \ang{90} to \ang{270}.  Panel 4 (Figure 5c) represents the end of the rotation from \ang{270} to \ang{360}.   By combining panels 1 through 4, a complete composite map representing the midpoint of CR 2099 is created.  The total open flux is then determined for CR 2099 by summing the calculated flux in each of these four panels.  This same procedure is implemented for each Carrington rotation in the dataset (CR 2056\,--\,2135).  This procedure of calculating open flux for each CR can be implemented without significant loss of accuracy since we are averaging over three Carrington rotations.\\
\tab \hspace{5mm}Figure \ref{AllResults-Plot} includes the total open flux deduced from coronal-hole observations in helium (paired with diachronic KPVT photospheric maps) and EUV (paired with synchronic ADAPT photospheric-field maps).  From the ADAPT-EUV result, two things are apparent.  First, the ADAPT-EUV open flux agrees closely with model-derived results from WSA.  The main disagreement between these results occurs at the beginning of 2012 where the model-derived open flux is briefly anti-correlated with the ADAPT-EUV result.  However, on the whole the ADAPT-EUV result tracks well with the WSA result.  The same is also true for the Harvey and Recely result using helium observations. Second, the total open flux derived with EUV observations and ADAPT maps is consistently lower than the spacecraft data especially around solar maximum.
%%%%%%%%%%%%%%%%%%%%%%%%%%%%%%%%%%%%%%%%%%%%%%%%%%%%%%%%%%%%%%%%%
%%%%%%%%%%%%%%%%%%%     Begin Section 4   %%%%%%%%%%%%%%%%%%%%%%%
%%%%%%%%%%%%%%%%%%   Discussion of Results  %%%%%%%%%%%%%%%%%%%%%
%%%%%%%%%%%%%%%%%%%%%%%%%%%%%%%%%%%%%%%%%%%%%%%%%%%%%%%%%%%%%%%%%    
\section{Discussion of Results} 
      \label{4-Results}           

\begin{figure}  
  \hspace{1mm}%
  % \centerline{\includegraphics[width=1.15\textwidth,scale=.5,trim={0cm 2.0cm 2.0cm 1.5cm},clip]{test_new_graph.pdf}} %%% To Trim: <left> <lower> <right> <upper>%%% To Trim: <left> <lower> <right> <upper>
    \centerline{\includegraphics[width=1\textwidth,trim={.4cm .7cm .4cm 5cm},clip]{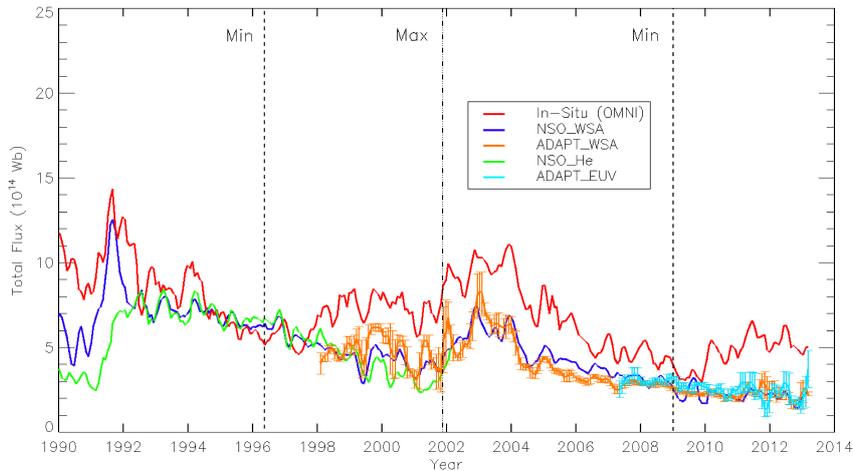}} %%% To Trim: <left> <lower> <right> <upper>
  %\centerline{\includegraphics[width=1.08\columnwidth,trim={0cm .9cm 0.5cm 24.0cm},clip]{open_flux_Figure_6.eps}} %%% To Trim: <left> <lower> <right> <upper>
 % \centerline{\includegraphics[width=1.15\textwidth,trim={0cm 3.5cm 0.2cm 4.0cm},clip]{open_flux_test.pdf}} %%% To Trim: <left> <lower> <right> <upper>
  
   % \centerline{\includegraphics[scale=.47,trim={0.5cm 2.0cm 2.0cm 1.5cm},clip]{open_flux_test.pdf}} %%% To Trim: <left> <lower> <right> <upper>%%% To Trim: <left> <lower> <right> <upper>
              \caption{Total open flux derived from WSA (using both diachronic and synchronic photospheric-field maps), {\it in-situ} observations, and manually contoured coronal holes in He and EUV, including vertical bars representing the standard deviation of the uncertainty in results using ADAPT photospheric-field maps.  All results are calculated over three-Carrington-rotation running averages.}
              \label{Open_flux_all_sdev-Plot}
   \end{figure}

The purpose of this work is to compare total unsigned open flux obtained using traditional methods ({\it i.e.} models and spacecraft single point measurements) with that derived from an alternative and independent approach ({\it i.e.} observationally derived coronal holes) to investigate the well-established issue of discrepancies between models and spacecraft observations of open flux. This study began with using both Carrington diachronic and ADAPT synchronic maps as input into WSA to see if using synchronic maps produced a better match between model-derived and spacecraft-observed open flux. Figure \ref{Figure1} shows the total open flux derived from WSA using both Carrington diachronic and ADAPT synchronic NSO/KPVT and VSM photospheric-field maps.  The two model-derived results agree rather well for the entire fifteen-year period of overlap.  This is not entirely surprising, since the two would likely vary from one another in detail from day to the next, but variations would be largely eliminated as apparent when averaging over three Carrington rotations. For the ADAPT-WSA result in this study, the standard deviation of the model uncertainty can be computed including 36 values for open flux ({\it i.e.} 12 ADAPT maps for each mid-rotation date of the three rotations).  Figure \ref{Open_flux_all_sdev-Plot} is the same as Figure \ref{AllResults-Plot}, now including vertical bars representing the range of variation of the ADAPT-WSA and ADAPT-EUV result.\\   
\tab \hspace{5mm}The range of ADAPT-WSA model variation is largest from $\approx$2000\,--\,2003 near Cycle 23 maximum, and less during periods of lower solar activity ($\approx$1998\,--\,1999 and 2004\,--\,2011).  There are also instances near Cycle 23 maximum where the range in the model result helps to close the gap between model and spacecraft derived open flux.
%%%%%%%%%%%%%%%%%%%%%%%%%%%%%%%%%%%%%%%%%%%%%%%%%%%%%%%%%%%%%%%
%%%%%%%%%%%%%%%%%%%%  Insert Figure 6  %%%%%%%%%%%%%%%%%%%%%%%%
%%%%%%%%%%%%%%%%%%%%%%%%%%%%%%%%%%%%%%%%%%%%%%%%%%%%%%%%%%%%%%%
%%%%%%%%%%%%%%%%%%%%%%%%%%%%%%%%%%%%%%%%%%%%%%%%%%%%%%%%%%%%%%%
%%%%%%%%%%%%%%%%%%%%  Insert Figure 6  %%%%%%%%%%%%%%%%%%%%%%%%
%%%%%%%%%%%%%%%%%%%%%%%%%%%%%%%%%%%%%%%%%%%%%%%%%%%%%%%%%%%%%%%
However, the results from Figure \ref{Open_flux_all_sdev-Plot} overall show that i) model-derived and spacecraft-observed open fluxes do not differ simply based on a consistent offset and ii) using synchronic photospheric maps as input into models does not account for the difference between models and {\it in-situ} observations.\\
\tab \hspace{5mm}Following the paradigm that the majority of open flux originates in coronal holes, observations of coronal holes can be paired with photospheric-field maps to calculate open flux as outlined in Section \ref{3-CHobs}.  To examine the usefulness of this method, Harvey and Recely's open-flux calculations from 1990 to 2003 were used in this study without modification (Figure \ref{AllResults-Plot}, green line). This data set was also derived with a set of diachronic KPVT photospheric-field maps that were identical to those used in the NSO-WSA result for this time period (Figure \ref{AllResults-Plot}, blue line).  Remarkably, total open flux derived using coronal holes manually identified in He track the WSA model-derived result almost exactly for almost nine years (1992\,--\,2000).\\
\tab \hspace{5mm}Given this overall agreement, we applied the same technique to more recent EUV observations of the corona to derive total open flux. This result (Figure \ref{AllResults-Plot}, cyan line) agrees well with WSA-derived results for the entire seven-year period (2007\,--\,2013) of overlap, with the exception of a brief anti-correlation at the beginning of 2012.  The previous discussion in Section \ref{3.1-He1083_Obs} regarding the difficulty of identifying mid-latitude coronal holes in helium during high solar activity as the potential source of low helium open flux estimates near Cycle 22 and 23 maximum (approx. 1990\,--\,1992 and 2000\,--\,2002) could be supported by the fact that we do not see lower EUV open flux estimates when compared to WSA from 2012\,--\,2013 approaching Cycle 24 maximum. Extending the ADAPT-EUV and model-derived result in a future study would be necessary to confirm this. Since EUV-derived coronal holes were paired with ADAPT maps to calculate the total open flux, we are again able to determine the range of variation in the photospheric field for 36 values of open flux over three Carrington rotations, shown in Figure \ref{Open_flux_all_sdev-Plot}.  Similar to the uncertainty range in the ADAPT-WSA result, variation is more pronounced during higher solar activity ($\approx$2011\,--\,2013), and the results vary little during solar minimum (2007\,--\,2010).\\
\tab \hspace{5mm}Since both results using observationally derived coronal holes to calculate open flux match well with WSA, and the same photospheric-field maps were used to obtain the radial magnetic field for each result, this suggests that WSA accurately reproduces the locations and areas of coronal holes identified in He and EUV observations at least on three-Carrington-rotation time scales.  Further, observations from spacecraft disagree with open flux derived from all other methods, except near solar minimum.  {\it In-situ} observations of open flux are consistently greater than all other estimates in this study from 1998 onward, with the most deviation occurring near solar maximum.\\ 
\tab \hspace{5mm}Our results agree with \citet{Lowder2017} and \citet{Linker2017} in that open-flux estimates from both models and coronal-hole observations are consistently lower than that obtained from spacecraft (\citealt{Lowder2017} Figure 4b, \citealp{Linker2017} Table 1 and 2). However, Lowder {\it et al.}'s EUV-derived open fluxes are regularly lower than those obtained with a PFSS model, where our WSA and EUV-derived open fluxes agree well on average.   In this case, it is difficult to compare since both methods of obtaining open flux in our study use the same set of photospheric-field maps whereas Lowder {\it et al.} used WSO maps to drive the PFSS model and SDO \textit{Helioseismic and Magnetic Imager} (HMI) maps to pair with STEREO and SDO EUV observations. In addition, a similar study to ours was conducted by \citet{Harvey2013} comparing radial magnetic-flux density derived from helium coronal holes and spacecraft observations, showing that the two agree well on average.  However, it is difficult to compare our results since Harvey averaged the measured $B_\text{r}$ over three days before taking the unsigned value, where we took $|B_\text{r}|$ from daily values.  Also, it is not mentioned what photospheric-field measurements were paired with He-derived coronal holes to calculated the flux density. \\ 
\tab \hspace{5mm}There are several possible explanations as to why both coronal-hole observations on average reproduce WSA-derived open flux and why spacecraft observations disagree with all other estimates.  One possibility is that there may be open-flux sources that are not traditionally defined as coronal holes and are therefore not dark in EUV. These sources could be due to time-dependent phenomena ({\it e.g.} opening/closing of magnetic-field lines near active regions) that are obscured in EUV emission and not captured by static-potential-field models.  Nonpotential fields such as dynamic active regions can also have a significant impact on model-derived coronal-hole boundaries \citep{MHDvsPFSS-Riley2006} and thus the amount of open flux. MHD models in case studies have generated larger open flux estimates when compared to PFSS models (see \citealp{Linker2017} Table 1); however, they still underestimate spacecraft measurements and also derive areas of open flux that are not observed in EUV as coronal holes. Both PFSS and MHD computations involve the tuning of free parameters that affect the estimated open flux. However, unlike PFSS solutions, MHD models describe the plasma properties of the corona and solar wind, including heating and acceleration.  A more comprehensive understanding of the physics behind such processes is necessary to provide more physical constraints to free parameters.  Observations from \textit{Parker Solar Probe} in the near future will help fulfill this need by providing {\it in-situ} measurements of the corona.\\
\tab \hspace{5mm}In this work, the source-surface height was fixed at the conventional value of 2.5 \(\text{R}_\odot\) for all WSA open-flux estimates.  Using this height produced excellent agreement between model-derived open flux and that obtained from He and EUV coronal-hole observations. We do not explore the effects of raising or lowering the source surface for different portions of the solar cycle for this reason. Studies such as \citealt{Lee2011} and \citealt{Arden2014} suggest that raising the source surface height by 15\,--\,30\,\% and 20\,\% respectively during solar minimum would reproduce the spacecraft-observed open flux from OMNI.   However, our results have the best agreement between model-derived open flux and OMNI data near solar minimum.  Since we do not make use of photospheric-field maps derived from the same sources as these studies we cannot make a direct comparison.\\
\tab \hspace{5mm}Second, inconsistencies and lack of global photospheric-field measurements likely contribute to model uncertainties in open-flux estimates.  Significant errors can be introduced when converting raw magnetograms into radial fields to make photospheric-field maps \citep{MHDvsPFSS-Riley2006}, and well-known offsets exist between magnetograms from different sources (\citealp{Riley2014}, Table 3). To mitigate the effects of these issues, both the synchronic ADAPT and diachronic Carrington maps for this work were created using magnetograms from the identical sources ({\it i.e.} NSO/KPVT and VSM data).  Likewise, the Sun's far side and polar regions are not directly observed. Averaging our results over three Carrington rotations helps reduce the effects of not having far-side observations, and using ADAPT helps to evolve flux where observations are poorly measured ({\it i.e.} poles) or have not recently been updated.  It is possible that there could be unaccounted for open flux at the poles that could create systematic underestimates  of open flux when using photospheric-field maps (\citealt{Linker2017}, \citealt{Tusenta2008}).  One would expect this to be an issue primarily near solar minimum, where we see large polar coronal holes in EUV. However, the largest difference that we find between spacecraft-observed open flux and all other methods is near solar maximum, but the agreement at solar minimum is certainly not exact.\\
\tab \hspace{5mm}One known issue with heliospheric observations is that {\it in-situ} spacecraft detect radial components of the magnetic field resulting from CMEs, which can still be closed for many days even in passing Earth (\citealt{Owens2008a}, \citealt{Riley2007}).  This problem is most severe near solar maximum, when the CME prediction rate is the greatest.  This is consistent with the discrepancies seen in our results in that spacecraft observations agree best with WSA near solar minimum. However, one might expect open flux obtained using EUV observations to account for CME flux if these field lines are dark in EUV \citep{Linker2017}, yet EUV-derived open flux does not produce that measured {\it in situ}. Also, kinematic effects likely produce an additional contribution to $B_\text{r}$ measurements as reviewed in Section~\ref{1-Intro}, specifically, large-scale solar-wind structures that tangentially flow into and distort other field lines.  These effects are corrected for in various ways (\citealp{Lockwood2009c}, \citealp{Owens2017}), yet we used daily  averaged $|B_\text{r}|$ values since averaging $B_\text{r}$ between 1 to 5 days before taking the unsigned value has given similar results to other correction methods on average (\citealp{Owens2017}, Figure 5).  It is worth further investigating to what extent kinematic effects alter $B_\text{r}$ measurements, and if applying correction method(s) to $B_\text{r}$  yields similar open flux values to both models and coronal-hole-derived flux.  

%%%%%%%%%%%%%%%%%%%%%%%%%%%%%%%%%%%%%%%%%%%%%%%%%%%%%%%%%%%%%%%%%
%%%%%%%%%%%%%%%%%%%     Begin Section 5   %%%%%%%%%%%%%%%%%%%%%%%
%%%%%%%%%%%%%%%%%%%%%%%%   Summary  %%%%%%%%%%%%%%%%%%%%%%%%%%%%%
%%%%%%%%%%%%%%%%%%%%%%%%%%%%%%%%%%%%%%%%%%%%%%%%%%%%%%%%%%%%%%%%%

\section{Summary} 
      \label{5-Summary} 

We have compared total unsigned open heliospheric flux derived from WSA, spacecraft single-point measurements, and pairing manually identified coronal holes from (diachronic) He {\sc I} 10830 \AA\ equivalent width and (synchronic) EUV maps with their respective type of photospheric-field maps.  The same set of KPVT/VSM diachronic and ADAPT KPVT/VSM synchronic photospheric-field maps were used with both WSA and observationally derived coronal holes to determine open flux. This study spanned 23 years (1990\,--\,2013) or two solar cycles, including all of Cycle 23 and several years of Cycles 22 and 24.  All of the open magnetic flux results were calculated over three-Carrington-rotation running averages. We find that the total open flux values obtained from He {\sc I} 10830 \AA\ and EUV observations agree quite well with that obtained from WSA, with the greatest disagreement occurring near solar maximum.  There is a period from $\approx$1994\,--\,1998 where open flux obtained from all three methods agrees on average, however, {\it in-situ} spacecraft-derived open flux estimates disagree with--and are larger than--those obtained from all other methods from 1998 onward.\\ 
\tab \hspace{5mm}Given the shortcomings of each method used in this study, it is likely that {\it{several}} factors contribute to the discrepancies between model and spacecraft-derived open flux.  While there are drawbacks with both static-potential-field models and the photospheric-field maps that drive them, the open-flux results obtained using WSA and coronal-hole observations on the whole agree exceptionally well with each other. There could be calibration issues with the photospheric-field maps; however, a simple offset does not explain the disagreement between open flux derived with WSA and spacecraft observations.  The source surface could also be lowered to open more flux and provide better agreement with {\it in-situ} measurements, yet the model-derived open flux would not match that derived from He {\sc I} 10830 \AA\ and EUV observations.\\
\tab \hspace{5mm}One possibility that could explain these discrepancies, which has not truly been addressed, is the temporal evolution of the magnetic field.  For example, steady-state models such as WSA could identify well magnetic fields that are continuously open, but miss those fields that are open intermittently.  If this is true, it would be problematic at coronal-hole boundaries and near active regions where field lines are constantly opening and closing, resulting in models under-predicting total open flux.  Further, under the assumption that open-field regions appear dark in EUV, one would expect EUV observations to capture the time-dependent evolution of the corona and produce open flux estimates that match {\it in-situ} measurements at least on average.  However, the open flux derived with EUV observations agrees better with that obtained from WSA.  Therefore, it is probable that there is unaccounted for open flux that cannot be resolved in steady-state models and does not originate from what are traditionally described as coronal holes (\textit{i.e.} low EUV emission).  As time-dependent models become more routine, it will then be possible to determine the contribution of intermittent open fields to the total open flux.  Finally, the possibility that spacecraft may overestimate total open flux should be further investigated.  Although this is certainly not the only source of discrepancy between model-derived and spacecraft-observed open flux, the potential impact of this effect should not be overlooked.\\

%%%%%%%%%%%%%%%%%%%%%%%%%%%%%%%%%%%%%%%%%%%%%%%%%%%%%%%%%%%%%%%%%%%%%%%%%%%
\begin{acks}
% * <demarcos13@gmail.com> 2018-06-14T15:25:11.210Z:
%
% ^.
 This work was partially supported by the Air Force Scholars Program.  We acknowledge use of NASA/GSFC's Space Physics Data Facility's OMNIWeb service and OMNI data.  This work utilizes ADAPT maps produced collaboratively between AFRL and NSO/NISP. NSO/Kitt Peak data used here are produced cooperatively by NSF/NSO, NASA/GSFC, and NOAA/SEL.   SOLIS data for this work are obtained and managed by NSO/NISP, operated by AURA, Inc. under a cooperative agreement with NSF.

\end{acks}

\section*{Disclosure of Potential Conflicts of Interest}  The authors declare that they have no conflicts of interest.

%\noindent To change a title use an optional parameter:\par
%\verb+\begin{acks}[Acknowledgements]...\end{acks}+

%\acknowledgment US spelling: \verb+\acknowledgment+
%\acknowledgement British  spelling: \verb+\acknowledgement+

%%%%%%%%%%%%%%%%%%%%%%%%%%%%%%%%%%%%%%%%%%%%%%%%%%%%%%%%%%%%%%%%%
%%%%%%%%%%%%%%%%%%%   Begin Bibliography   %%%%%%%%%%%%%%%%%%%%%%
%%%%%%%%%%%%%%%%%%%%%%%%%%%%%%%%%%%%%%%%%%%%%%%%%%%%%%%%%%%%%%%%%

% format of references provided by the journal (.bst)
\bibliographystyle{spr-mp-sola}
% name your Bibtex file containing your references (.bib)
\bibliography{sola_bibliography_example}  

% Checking: look if the file containing the ``\bibitem'' exits
%           so check if the .bbl file exist (bibTeX compilation)
\IfFileExists{\jobname.bbl}{} {\typeout{}
\typeout{****************************************************}
\typeout{****************************************************}
\typeout{** Please run "bibtex \jobname" to obtain} \typeout{**
the bibliography and then re-run LaTeX} \typeout{** twice to fix
the references !}
\typeout{****************************************************}
\typeout{****************************************************}
\typeout{}}

\end{article} 

\end{document}